\documentclass{article}
\PassOptionsToPackage{numbers,sort&compress}{natbib}
\usepackage[preprint]{neurips_2026}

\usepackage[utf8]{inputenc}
\usepackage[T1]{fontenc}
\usepackage[english]{babel}
\usepackage{amsmath,amssymb}
\usepackage{graphicx}
\usepackage{booktabs}
\usepackage{microtype}
\usepackage{xcolor}
\usepackage{array}
\usepackage{placeins}
\usepackage{enumitem}
\usepackage{listings}
\usepackage[most]{tcolorbox}
\usepackage{hyperref}
\hypersetup{pdftitle={Coupled Dynamical System for Sparsemax Attention},pdfauthor={Lianlong Sun, Michael Huang}}

\definecolor{codebg}{HTML}{F7F8FA}
\definecolor{codeframe}{HTML}{C9CED6}
\definecolor{codekw}{HTML}{0B5CAD}
\definecolor{codestr}{HTML}{A33A1F}
\definecolor{codefn}{HTML}{6F42C1}
\definecolor{codecmt}{HTML}{6A737D}
\lstdefinestyle{python}{
  language=Python,
  basicstyle=\ttfamily\footnotesize,
  keywordstyle=\color{codekw}\bfseries,
  stringstyle=\color{codestr},
  commentstyle=\color{codecmt}\itshape,
  morekeywords={as,assert},
  deletekeywords={print,sum},
  deletekeywords={[2]print,sum},
  emph={ising,kuramoto,solve_ivp},
  emphstyle=\color{codefn},
  columns=fullflexible,
  keepspaces=true,
  showstringspaces=false,
  upquote=true,
}
\newtcblisting[auto counter]{pythoncode}[2][]{
  listing only,
  listing options={style=python},
  enhanced,
  colback=codebg,
  colframe=codeframe,
  colbacktitle=codeframe!45,
  coltitle=black,
  fonttitle=\small,
  title={Listing~\thetcbcounter: #2},
  boxrule=0.5pt,
  arc=2pt,
  left=6pt, right=6pt, top=2pt, bottom=2pt,
  #1,
}

\title{Coupled Dynamical System for Sparsemax Attention}
\author{%
  Lianlong Sun\thanks{Corresponding author.} \\
  University of Rochester \\
  \texttt{lianlongsun@rochester.edu}
  \And
  Michael Huang \\
  University of Rochester \\
  \texttt{michael.huang@rochester.edu}
}

\begin{document}
\maketitle

\begin{abstract}
We present a family of coupled dynamical systems that compute sparsemax attention. Each system fuses query-key similarity and normalization into a single physical relaxation. Specifically, we cast sparsemax attention as a constrained energy minimization problem and derive a pair of coupled differential equations: one drives the energy minimization and the other enforces the constraint. At equilibrium, the system state gives exactly the sparsemax attention probabilities. We realize this formulation with continuous Ising spins and with Kuramoto oscillators. These relaxations run natively on emerging non-von Neumann platforms such as Ising machines, which have demonstrated orders-of-magnitude higher energy efficiency than digital processors, yet have so far been used mostly for unconstrained quadratic combinatorial optimization. Because analog hardware has limited precision and is noisy, we simulate both systems under these non-idealities, and inference quality degrades gracefully. Our formulation extends these emerging computing substrates from combinatorial optimization to attention, opening a path toward a new class of physical accelerators for transformer inference.
\end{abstract}

\section{Introduction}
\label{sec:intro}

The transformer is the dominant architecture of modern deep learning~\citep{vaswani2017attention}. It was introduced for machine translation and now underlies large language models~\citep{radford2018improving,radford2019language} and vision models~\citep{dosovitskiy2021image,touvron2021deit}. Its central component is attention, which lets each token gather information directly from the other tokens of a sequence. For each token, attention compares its query with the keys of the other tokens. It turns the resulting scores into attention probabilities and returns the average of the values, taken with these probabilities. A transformer repeats this operation in many heads and layers, and its cost grows quadratically with the sequence length. The normalization that turns scores into probabilities shapes how attention behaves. The standard choice is softmax, which gives every key a nonzero probability. Sparse normalizers such as sparsemax~\citep{martins2016sparsemax} and entmax~\citep{peters2019sparse} instead give exactly zero probability to low-scoring keys. They keep the probabilities summing to one and focus each query on fewer keys, which makes attention easier to interpret~\citep{martins2016sparsemax,correia2019adaptively}.

Ising machines are physical systems that compute by natural relaxation~\citep{mohseni2022ising,zhang2024reviewising}. Their spins interact through programmable couplers and settle into a state of low Ising energy, which encodes the solution~\citep{lucas2014ising}. Ising machines have been built on many substrates. These include quantum annealers~\citep{johnson2011quantum}, coherent optical machines~\citep{inagaki2016coherent,mcmahon2016fully,yamamoto2017coherent,honjo2021100}, coupled oscillators~\citep{wang2019oim,chou2019analog,moy2022ring,vaidya2022creating,cilasun2025coupled,sun2025general}, and CMOS circuits~\citep{yamaoka2016ising,afoakwa2021brim,su2023reconfigurable,liu2025integrated}. They have demonstrated orders-of-magnitude higher energy efficiency than digital processors~\citep{afoakwa2021brim,cilasun2025coupled,zhang2024reviewising}. Their applications include Max-Cut~\citep{mcmahon2016fully,haribara2016coherent,wang2019oim,bashar2020experimental}, Boolean satisfiability~\citep{bashar2023designing,bybee2023efficient,sharma2023augmenting,sun2025general} and MIMO detection~\citep{singh2022ising,sreedhara2023mu}. Other applications include portfolio optimization, vehicle routing, clustering, chemical design and power systems~\citep{tanahashi2019application,parizy2022cardinality,bao2023ising,matsumoto2022distance,mao2023chemical,kirihara2023exploring}. Recent work also trains Boltzmann networks~\citep{niazi2024training}, learns graphs and functions~\citep{song2024ds,wu2024extending,liu2025instatrain,liu2025expressive}, and generates images~\citep{sun2026conditional}. Yet almost all of these applications share a common principle. At the model level, the machine runs the same simple dynamics: each spin is driven by its couplings to the other spins and a bias. A quadratic unconstrained binary optimization (QUBO) problem~\citep{lucas2014ising} can be naturally mapped to the model. Attention does not fit this form. A few works use Ising models to sample or search discrete attention patterns~\citep{liu2026ising,du2025qama,kim2026boltzmann}. None of them lets a physical system relax to the attention probabilities themselves.

We revisit where the efficiency of an Ising machine comes from. It computes by physical relaxation on a few simple analog primitives, such as spins, couplers and bias currents. We change the dynamics while keeping these primitives. The new dynamics must meet two goals. First, they must go beyond the simple Ising dynamics and have the expressivity to compute transformer attention. Second, they must map naturally onto Ising machines or similar analog computing substrates. Therefore, existing hardware designs can be largely reused, and the efficiency gain is preserved. The result is no longer the simple Ising dynamics. We call it a coupled dynamical system rather than an Ising machine. It couples the spin dynamics with a second dynamics that enforces a constraint.

Sparsemax attention is a natural target for this extension. Sparsemax is the solution of a quadratic program~\citep{martins2016sparsemax}. Its objective is a quadratic energy, as in an Ising machine. The new challenge is the constraint that the probabilities sum to one. From the Lagrangian of this program, we derive a pair of coupled ordinary differential equations (ODEs). The first mainly lowers the energy. Each key relaxes under its own query-key score and one shared feedback signal. The second mainly enforces the constraint. It compares the sum of the probabilities with one and adjusts the feedback signal. At equilibrium, the state gives exactly the sparsemax attention probabilities. Therefore, query-key similarity and normalization are fused into a single physical relaxation, with potential efficiency gains over conventional digital computing.

We realize this formulation with continuous Ising spins and with Kuramoto oscillators~\citep{kuramoto1975self}. Each key is one spin or one oscillator, and its couplings to the query form the scores. Every term maps onto a primitive that Ising machine hardware already provides~\citep{afoakwa2021brim,wang2019oim,lo2023ising}. A constraint node runs the second ODE. At the hardware level, Ising machines usually compute in analog circuits, without analog-to-digital conversion during relaxation. Analog substrates suffer from limited precision and noise~\citep{xiao2022accuracy}. We therefore simulate both systems under quantization and read noise to measure their inference quality. This serves as an evaluation of the coupled dynamical systems.

Our contributions are as follows.
\begin{itemize}[leftmargin=2em]
  \item We present a family of coupled dynamical systems that extends the simple Ising dynamics. We derive a pair of coupled ODEs whose equilibrium is exactly sparsemax attention.
  \item We map the dynamics onto Ising machine primitives, with continuous Ising spins and with Kuramoto oscillators. This mapping serves as a recipe for building attention on Ising machines and similar analog substrates, with potential efficiency gains over conventional digital computing.
  \item We simulate both systems inside transformers and evaluate their inference quality under quantization and read noise. The results show that the quality of both systems degrades gracefully. This provides the foundation for future hardware implementations.
\end{itemize}

\section{Methods}
\label{sec:method}

Sparsemax can be formulated as the solution of a quadratic program~\citep{martins2016sparsemax}. We first turn this program into a pair of coupled ODEs. We then map each term onto an Ising machine primitive, with continuous Ising spins and with Kuramoto oscillators. Appendix~\ref{app:code} gives a minimal implementation of both.

\subsection{From sparsemax to coupled ODEs}
\label{sec:method-sparsemax}

Let $q\in[-1,1]^d$ be a query, and let the $n$ keys form the rows of $K\in\mathbb{R}^{n\times d}$. The score of key $i$ is $s_i=\sum_jK_{ij}q_j/\sqrt d$. Sparsemax maps the scores to the closest point of the probability simplex~\citep{martins2016sparsemax}:
\begin{equation}
  \operatorname{sparsemax}(s)=\arg\min_{a_i\geq0,\ \sum_ia_i=1}\ \tfrac12\|a-s\|^2=[s-\tau]_+.
  \label{eq:sparsemax}
\end{equation}
Here $a$ is the vector of attention probabilities, with one entry $a_i$ per key. The bracket is $[\cdot]_+=\max(\cdot,0)$, and the threshold $\tau$ makes the output sum to one. Keys with $s_i\leq\tau$ get zero probability. We add a Lagrange multiplier $\mu$ for the sum-to-one constraint and keep each probability $a_i$ in $[0,1]$. This gives the Lagrangian
\begin{equation}
  \mathcal{L}(a,\mu)=\tfrac12\|a-s\|^2+\mu\Big(\sum_ia_i-1\Big),
  \qquad 0\leq a_i\leq1.
  \label{eq:lagrangian}
\end{equation}
Descent in $a$ and ascent in $\mu$ give our pair of coupled ODEs:
\begin{equation}
  \frac{da_i}{dt}=s_i-a_i-\mu\quad\text{with } a_i\in[0,1],
  \qquad\qquad
  \frac{d\mu}{dt}=\sum_ia_i-1.
  \label{eq:ode}
\end{equation}
Here $t$ is time. The first equation mainly lowers the energy $\tfrac12\|a-s\|^2$. Each key needs only its own score and the shared feedback signal $\mu$. The second equation mainly enforces the constraint that the probabilities sum to one. One node sums the probabilities and moves $\mu$ until the sum is one. We call it the constraint node. At equilibrium, each probability is $s_i-\mu$ clipped to $[0,1]$, and the probabilities sum to one. This is exactly sparsemax, with $\mu$ in the role of $\tau$.

\subsection{Continuous Ising spins}
\label{sec:method-ising}

We start from the linear Ising machine. At the model level, each spin $i$ has a continuous state $x_i$ and follows linear dynamics:
\begin{equation}
  \frac{dx_i}{dt}=\sum_jJ_{ij}x_j+h_i,
  \qquad
  x_i\in[-1,1].
  \label{eq:ising-textbook}
\end{equation}
Here $J_{ij}$ is the coupling from spin $j$ to spin $i$, and $h_i$ is the bias of spin $i$. The bound $[-1,1]$ is part of the model. At the hardware level, $x_i$ is the voltage on a spin capacitor, and couplers and bias currents charge it~\citep{afoakwa2021brim,liu2025integrated}. The bound is then the voltage limit of the circuit.

\paragraph{Mapping.} We use two groups of spins. The $d$ query spins hold the input, and query spin $j$ is clamped at $q_j$. The $n$ key spins hold the output, and key spin $i$ gives the probability $a_i=(x_i+1)/2$. This readout maps the spin range $[-1,1]$ onto the probability range $[0,1]$. We program the machine with
\begin{equation}
  J_{ij}=\frac{2K_{ij}}{\sqrt d},
  \qquad
  J_{ii}=-1,
  \qquad
  h_i=-1-2\mu.
  \label{eq:ising-map}
\end{equation}
The coupling $J_{ij}$ from query spin $j$ to key spin $i$ stores the key entry $K_{ij}$. The self-coupling $J_{ii}$ pulls each key spin back toward its input. The bias $h_i$ is the same for all key spins, and the constraint node sets it through $\mu$. Key spins do not couple to each other.

\paragraph{Dynamics.} With these settings and the constraint node, the machine runs
\begin{equation}
  \frac{dx_i}{dt}=\underbrace{\sum_jJ_{ij}q_j}_{2s_i}-x_i-1-2\mu,
  \qquad
  \frac{d\mu}{dt}=\sum_i\frac{x_i+1}{2}-1 .
  \label{eq:ising}
\end{equation}
With $a_i=(x_i+1)/2$, these are exactly the coupled ODEs of Eq.~\eqref{eq:ode}. The key spins compete through the shared bias. When their probabilities sum to more than one, $\mu$ rises and lowers every key spin. A key spin with $s_i\leq\mu$ settles at the bound $x_i=-1$ and gets exactly zero probability. The other key spins settle inside the bound and carry the nonzero probabilities. 

\subsection{Kuramoto oscillators}
\label{sec:method-kuramoto}

An oscillator-based Ising machine (OIM) stores each spin in the phase $\theta_i$ of an oscillator~\citep{wang2019oim}. At the model level, the phases follow Kuramoto dynamics~\citep{kuramoto1975self} with second-harmonic injection:
\begin{equation}
  \frac{d\theta_i}{dt}=-\sum_jJ_{ij}\sin(\theta_i-\theta_j)-K_s\sin2\theta_i .
  \label{eq:oim}
\end{equation}
Here $J_{ij}$ is the coupling between oscillators $i$ and $j$, and $K_s$ is the strength of the second-harmonic injection. At the hardware level, couplers link the oscillators, and a second-harmonic signal is injected into each of them. We reuse these primitives.

\paragraph{Mapping.} We use three kinds of oscillators. Each key is one oscillator with phase $\theta_i$. Each query component is held by a pair of oscillators with fixed phases $\pm\varphi_j$. A reference oscillator has fixed phase $0$. A pair acts on a key oscillator as one input of strength $2\cos\varphi_j$, because $\sin(\theta-\varphi)+\sin(\theta+\varphi)=2\cos\varphi\sin\theta$. We program the machine with
\begin{equation}
  \cos\varphi_j=q_j,
  \qquad
  J_{ij}=\frac{K_{ij}}{4\sqrt d},
  \qquad
  J_{i0}=-\mu,
  \qquad
  K_s=-\frac12 .
  \label{eq:kuramoto-map}
\end{equation}
The coupling $J_{ij}$ links key oscillator $i$ to both oscillators of pair $j$ and stores the key entry $K_{ij}$. The coupling $J_{i0}$ links key oscillator $i$ to the reference, and the constraint node sets it through $\mu$.

\paragraph{Dynamics.} With these settings, the key oscillators run
\begin{equation}
  \frac{d\theta_i}{dt}=-\sum_jJ_{ij}\Big[\sin(\theta_i-\varphi_j)+\sin(\theta_i+\varphi_j)\Big]+\mu\sin\theta_i+\frac12\sin2\theta_i .
  \label{eq:kuramoto}
\end{equation}
This is the OIM dynamics of Eq.~\eqref{eq:oim} with the settings of Eq.~\eqref{eq:kuramoto-map}. The couplings to the query pairs deliver $-\tfrac{s_i}{2}\sin\theta_i$ to key oscillator $i$. The readout is the rectified cosine of the phase. The constraint node compares the sum of the readouts with one half:
\begin{equation}
  r_i=[\cos\theta_i]_+,
  \qquad
  a_i=2r_i,
  \qquad
  \frac{d\mu}{dt}=\sum_ir_i-\frac12 .
  \label{eq:kuramoto-readout}
\end{equation}
As with the spins, the key oscillators compete through the shared reference coupling. At equilibrium, $a_i=[s_i-2\mu]_+$ and the probabilities sum to one. This is exactly sparsemax, with $\tau=2\mu$.

\subsection{More realizations}
\label{sec:method-general}

Continuous Ising spins and Kuramoto oscillators are two realizations of one principle. The principle is not tied to these two designs. The pair of coupled ODEs in Eq.~\eqref{eq:ode} specifies what a substrate computes for attention. Each Ising machine or analog platform can realize it with its own primitives, readout and scaling. It can also reshape the dynamics to fit its physics.

\section{Evaluation}
\label{sec:eval}

With ideal hardware, both systems give the same predictions as a digital implementation of sparsemax. Analog substrates, however, suffer from non-idealities such as limited precision, noise, and device mismatch~\citep{xiao2022accuracy}. We therefore study the inference quality under quantization and read noise.

\subsection{Setup}
\label{sec:eval-setup}

\paragraph{Language model.} We train a GPT~\citep{radford2019language} with 51.5M parameters from scratch on FineWeb-Edu~\citep{penedo2024fineweb}. It has 8 layers with 8 heads each, width 512, the GPT-2 vocabulary and a context of 1024 tokens. It trains on 0.8B tokens with AdamW~\citep{loshchilov2019decoupled}. We report the perplexity on 512 held-out windows of 1024 tokens and the zero-shot LAMBADA accuracy~\citep{paperno2016lambada} from the LM Evaluation Harness~\citep{gao2024harness}.

\paragraph{Vision transformer.} We start from DeiT-S/16~\citep{touvron2021deit,dosovitskiy2021image}, pretrained on ImageNet-1k~\citep{deng2009imagenet,russakovsky2015imagenet}. We fine-tune it for 15 epochs on the 100-class subset ImageNet-100~\citep{tian2020cmc}. During the first 3 epochs, $\alpha$-entmax~\citep{peters2019sparse} ramps from $\alpha=1$ to $2$, and then sparsemax is used. We report top-1 accuracy on the 5,000 validation images.

\paragraph{Bounded query.} Query spins are bounded to $[-1,1]$ for convenience, and $\cos\varphi_j$ of a query oscillator pair lies in the same range (Sections~\ref{sec:method-ising} and~\ref{sec:method-kuramoto}). We therefore apply $\tanh$ to the query projection of each head and call this model \emph{bounded sparsemax}. All evaluated models use bounded sparsemax in every head.

\paragraph{Non-ideal hardware.} At inference, a coupled dynamical system computes the attention probabilities of every head in every layer. The value vectors are then averaged with these probabilities digitally. We simulate each relaxation for $T=20t_0$, where $t_0$ is the time constant of a spin or an oscillator. We use an adaptive Heun-Euler method. To model non-ideal hardware, we perturb the query $q$ and the keys $K$ programmed into the system. Both are quantized to a fixed number of bits, and Gaussian read noise is added with a level relative to their full range. The models are not retrained for different non-ideal cases. With read noise, we report the mean and standard deviation over 20 noise draws.

\subsection{Inference quality under non-ideal conditions}
\label{sec:eval-results}

\begin{table}[t]
  \centering
  \caption{Inference quality with a coupled dynamical system in every attention head.
  The first row is ideal hardware and equals exact sparsemax.
  The other rows quantize $q$ and $K$, add read noise, or both.
  With read noise, entries are the mean $\pm$ standard deviation over 20 noise draws.}
  \label{tab:inference}
  \footnotesize
  \setlength{\tabcolsep}{3pt}
  \newcommand{\sd}[1]{\,{\scriptsize$\pm$#1}}
  \begin{tabular}{llcccccc}
    \toprule
    & & \multicolumn{2}{c}{ImageNet-100 top-1 (\%) $\uparrow$} & \multicolumn{2}{c}{FineWeb-Edu perplexity $\downarrow$} & \multicolumn{2}{c}{LAMBADA acc. (\%) $\uparrow$} \\
    \cmidrule(lr){3-4}\cmidrule(lr){5-6}\cmidrule(lr){7-8}
    Precision & Read noise & Ising & Kuramoto & Ising & Kuramoto & Ising & Kuramoto \\
    \midrule
    Full & None & 91.02 & 91.02 & 35.60 & 35.60 & 13.02 & 13.02 \\
    Full & 2\% & 90.77\sd{0.12} & 90.77\sd{0.12} & 35.90\sd{0.01} & 35.90\sd{0.01} & 12.94\sd{0.13} & 12.94\sd{0.13} \\
    Full & 5\% & 90.31\sd{0.23} & 90.31\sd{0.23} & 37.78\sd{0.02} & 37.78\sd{0.02} & 11.97\sd{0.21} & 11.97\sd{0.21} \\
    \midrule
    5-bit & None & 90.74 & 90.86 & 35.87 & 35.87 & 12.89 & 12.87 \\
    5-bit & 2\% & 90.63\sd{0.16} & 90.65\sd{0.18} & 36.19\sd{0.01} & 36.19\sd{0.01} & 12.65\sd{0.12} & 12.64\sd{0.11} \\
    5-bit & 5\% & 90.22\sd{0.19} & 90.24\sd{0.20} & 38.14\sd{0.02} & 38.14\sd{0.02} & 11.77\sd{0.22} & 11.78\sd{0.22} \\
    \midrule
    4-bit & None & 90.54 & 90.50 & 37.03 & 37.03 & 12.25 & 12.25 \\
    4-bit & 2\% & 90.50\sd{0.19} & 90.46\sd{0.18} & 37.41\sd{0.01} & 37.41\sd{0.01} & 12.10\sd{0.16} & 12.10\sd{0.16} \\
    4-bit & 5\% & 89.95\sd{0.19} & 89.97\sd{0.21} & 39.64\sd{0.03} & 39.64\sd{0.03} & 11.35\sd{0.22} & 11.34\sd{0.22} \\
    \bottomrule
  \end{tabular}
\end{table}

Table~\ref{tab:inference} shows the results. With ideal hardware, both systems give the predictions of exact sparsemax on every image and every LAMBADA example. We do not apply quantization-aware or noise-aware training. Quality degrades gracefully with lower precision and higher noise. At 5 bits with 2\% noise, mean top-1 accuracy drops by 0.4 points and perplexity rises by 1.7\%. The worst case is 4 bits with 5\% noise. It costs 1.1 points of top-1 accuracy and 1.7 points on LAMBADA, and it raises perplexity by 11\%. Across all cases, Ising spins and Kuramoto oscillators deliver similar performance.

\section{Related work}
\label{sec:related}

\paragraph{Ising machines and oscillator networks.} Ising machines program a problem into couplings and biases and let a physical system relax to a low-energy state~\citep{mohseni2022ising,zhang2024reviewising}. They have been built with superconducting qubits~\citep{johnson2011quantum}, optical parametric oscillators~\citep{inagaki2016coherent,inagaki2016large,mcmahon2016fully,yamamoto2017coherent,hamerly2019experimental,honjo2021100}, probabilistic bits~\citep{borders2019integer,aadit2022massively}, CMOS circuits~\citep{afoakwa2021brim,su2023reconfigurable,wang2024energy,liu2025integrated,liu2025ising}, Josephson parametric oscillators~\citep{razmkhah2024josephson,ucpinar2024scalable} and coupled electronic oscillators~\citep{wang2019oim,chou2019analog,vaidya2022creating,moy2022ring,albertsson2023highly,lo2023ising,cilasun2025coupled}. Oscillator-based machines follow Kuramoto dynamics, whose computational properties are well studied~\citep{kuramoto1975self,erementchouk2022computational,ekanayake2025different}. Ising machines solve Max-Cut, satisfiability, MIMO detection and other optimization problems~\citep{mcmahon2016fully,haribara2016coherent,wang2019oim,bashar2020experimental,ucpinar2024scalable,shaglel2026comprehensive,bashar2023designing,bybee2023efficient,sharma2023augmenting,su2023reconfigurable,sikhakollu2024high,sun2025general,singh2022ising,sreedhara2023mu,singh2024uplink,tanahashi2019application,matsumoto2022distance,parizy2022cardinality,bao2023ising,kirihara2023exploring,mao2023chemical,tsuyumine2024optimization}. Several works extend the simple Ising dynamics with higher-order interactions~\citep{bybee2023efficient,sharma2023augmenting,sun2025general,de2025incorporate} or with one error variable per spin~\citep{leleu2019destabilization,leleu2021scaling}. Beyond optimization, Ising machines and related dynamical systems learn, solve differential equations and generate images~\citep{niazi2024training,laydevant2024training,wang2024training,song2024ds,wu2024extending,liu2025instatrain,liu2025expressive,liu2025ds,sun2026conditional}. We extend this line of work by changing the Ising dynamics itself while keeping it hardware friendly. This brings Ising machines to modern machine learning tasks such as transformer attention.

\paragraph{Attention.} Attention was introduced to let a translation model align each output word with the relevant input words~\citep{bahdanau2015neural}. The transformer builds on scaled dot-product attention, with a softmax that normalizes the scores of each query~\citep{vaswani2017attention}. Because its cost grows quadratically with the sequence length, many works make attention cheaper. Fixed sparse patterns restrict in advance which query-key pairs are scored~\citep{child2019generating,beltagy2020longformer,zaheer2020big}. Linear attention replaces the softmax kernel by feature maps and reaches linear cost~\citep{katharopoulos2020transformers,choromanski2021rethinking}. FlashAttention keeps exact softmax attention and reorganizes its memory accesses on GPUs~\citep{dao2022flashattention}. Element-wise ReLU or sigmoid attention drops the sum-to-one constraint~\citep{zhang2021sparse,wortsman2023replacing,ramapuram2025theory}. Sparsemax and entmax keep the constraint and give low-scoring keys exactly zero probability~\citep{martins2016sparsemax,peters2019sparse,correia2019adaptively}. Softmax attention provably disperses as the input grows, and sparse normalizers avoid this~\citep{velickovic2025softmax,vasylenko2026long}. Our design fuses the query-key scores and the normalization into one physical relaxation, whose equilibrium is exactly sparsemax attention.

\paragraph{Attention as a dynamical system or energy minimization.} Hopfield networks retrieve patterns by lowering an energy~\citep{hopfield1984neurons}. In modern Hopfield networks, one update step on this energy is softmax attention~\citep{ramsauer2021hopfield}, and sparse variants give sparsemax and entmax attention~\citep{hu2023sparse,santos2024sparse}. The Energy Transformer replaces the stack of blocks by one recurrent block that descends an engineered energy over the tokens~\citep{hoover2023energy}. Optimization views read self-attention as a descent step on a token energy or coding rate~\citep{yang2022transformers,yu2023white}. Sinkformers replace the softmax by Sinkhorn iterations, which make the attention matrix doubly stochastic~\citep{sander2022sinkformers}. Other works model tokens across layers as interacting particles that cluster over time~\citep{lu2019understanding,geshkovski2023emergence,geshkovski2025mathematical}. Optimization layers output the solution of a quadratic program~\citep{amos2017optnet}, and classic neural networks enforce constraints exactly through Lagrange multiplier dynamics~\citep{platt1987constrained,zhang1992lagrange}. AKOrN replaces threshold units by Kuramoto oscillators~\citep{miyato2025akorn}. Kuramoto Attention and KoPE give each token phases that synchronize through softmax-normalized couplings~\citep{nunley2026kuramoto,xiao2026kuramoto}. We also compute attention as the equilibrium of a dynamical system. Our dynamics act on the attention probabilities themselves and map onto Ising machine primitives.

\section{Conclusion}
\label{sec:conclusion}

We presented a family of coupled dynamical systems that extends the Ising dynamics to transformer attention. From the sparsemax quadratic program, we derived a pair of coupled ODEs whose equilibrium is exactly sparsemax attention. Both ODEs map term by term onto Ising machine primitives, with continuous Ising spins and with Kuramoto oscillators. This mapping is a recipe for building attention on Ising machines and similar analog substrates. We evaluated both systems inside a language model and a vision transformer. Their inference quality degrades gracefully with lower precision and higher read noise, which paves the way for larger-scale work.

Scaling these systems to a full transformer opens problems at the architecture, circuit and algorithm levels. At the architecture level, attention must integrate with the projections and feed-forward layers, and the system must scale to long contexts. At the circuit level, the constraint node requires a new circuit that senses all spins and drives one shared bias back to every spin. At the algorithm level, the key-value cache, the attention heads and the training of the models can be co-designed with the hardware. Solving these problems leads from dynamical systems to physical accelerators for efficient transformer inference.

\newpage

\bibliographystyle{unsrtnat}
\bibliography{main}

\newpage
\appendix
\section{Implementation}
\label{app:code}

Listing~\ref{lst:code} reproduces both systems on an example. The function \texttt{ising} is Eq.~\eqref{eq:ising}. The function \texttt{kuramoto} is Eq.~\eqref{eq:oim} with the settings of Eq.~\eqref{eq:kuramoto-map} and the readout of Eq.~\eqref{eq:kuramoto-readout}.

\begin{pythoncode}[label=lst:code]{Minimal reproduction of both coupled dynamical systems.}
import numpy as np
from scipy.integrate import solve_ivp

q = np.array([0.6, 0.8])
K = np.array([[1., 1.], [1., 0.5], [-1., 1.]])
V = np.array([[1., 0.], [0., 1.], [-1., -1.]])
n, d = K.shape

s = K @ q / np.sqrt(d)
z = np.sort(s)[::-1]
cutoffs = (np.cumsum(z) - 1) / np.arange(1, n + 1)
a = np.maximum(s - cutoffs[z > cutoffs][-1], 0)

def ising(t, state):
    x, mu = np.clip(state[:-1], -1, 1), state[-1]
    J = 2 * K / np.sqrt(d)
    dx = J @ q - x - 1 - 2 * mu
    dx[((x <= -1) & (dx < 0)) | ((x >= 1) & (dx > 0))] = 0
    return np.r_[dx, np.sum((x + 1) / 2) - 1]

def kuramoto(t, state):
    theta, mu = state[:-1], state[-1]
    phi = np.arccos(q)
    J = K / (4 * np.sqrt(d))
    th = theta[:, None]
    pairs = (J * (np.sin(th - phi) + np.sin(th + phi))).sum(1)
    dtheta = -pairs + mu * np.sin(theta) + 0.5 * np.sin(2 * theta)
    return np.r_[dtheta, np.sum(np.maximum(np.cos(theta), 0)) - 0.5]

print(f"{'Sparsemax':9}: {a} -> {a @ V}")
for name, ode, x0, readout in [
    ("Ising", ising, np.full(n, 2 / n - 1),
     lambda x: (np.clip(x, -1, 1) + 1) / 2),
    ("Kuramoto", kuramoto, np.full(n, np.pi / 2),
     lambda theta: 2 * np.maximum(np.cos(theta), 0)),
]:
    sol = solve_ivp(ode, (0, 200), np.r_[x0, 0.], rtol=1e-9, atol=1e-11)
    b = readout(sol.y[:-1, -1])
    print(f"{name:9}: {b} -> {b @ V}")
    assert sol.success and np.allclose(a, b, atol=1e-6, rtol=0)
\end{pythoncode}
\end{document}